\documentstyle[12pt]{article}
\title{\bf Curvature Expansion\\ 
for the Gluodynamics String\\
including Perturbative Gluonic Contributions.}
\author {D.V.ANTONOV\thanks{E-mail addresses: 
antonov@vxitep.itep.ru, antonov@pha2.physik.hu-berlin.de}{\,}
\thanks {Supported
by Graduiertenkolleg {\it Elementarteilchenphysik}, Russian
Fundamental Research Foundation, Grant No.96-02-19184, DFG-RFFI,
Grant 436 RUS 113/309/0 and by the Intas, Grant No.94-2851.}\\
{\it Institute of Theoretical and Experimental Physics,}\\
{\it B.Cheremushkinskaya 25, 117 218, Moscow, Russia}\\
{\it and}\\
{\it Institut f\"ur Physik, Humboldt-Universit\"at, Berlin}\\
\\ and\\
\\
D.EBERT \thanks{E-mail address: debert@qft2.physik.hu-berlin.de}\\
{\it Institut f\"ur Physik, 
Humboldt-Universit\"at,}\\
{\it Invalidenstrasse 110, D-10115, Berlin, Germany}\\}
\date{}
\begin{document}
\maketitle
\vspace{1mm}
\centerline{\bf {Abstract}}
\vspace{3mm}
Perturbation theory in the nonperturbative QCD vacuum and the
non-Abelian Stokes theorem, representing a Wilson loop in the $SU(2)$
gluodynamics as an integral over all the orientations in colour
space, are applied to a derivation of the correction to the string
effective action in the lowest order in the coupling constant $g$. 
This correction is due to the 
interaction of perturbative gluons with the string world sheet and 
affects only the coupling constant of
the rigidity term, while its contribution to the string tension of
the Nambu-Goto term vanishes. The obtained correction to the rigidity 
coupling constant multiplicatively depends on the colour ``spin'' of the
representation of the Wilson loop under consideration and a certain 
path integral, which includes the 
background Wilson loop average.

\vspace{6mm}
{\large \bf 1. Introduction}

\vspace{3mm}
Recently, a new approach to the gluodynamics string was
suggested$^{1}$. Within this approach, one considers the Wilson loop 
average written through the non-Abelian Stokes theorem$^{2,3}$ and
the cumulant expansion$^{3,4}$ as a statistical weight in the
partition function of some effective string theory. The action of
this string theory may then be obtained via expansion of the averaged 
Wilson loop in powers of two small parameters,
$\left(\alpha_s\left<F_{\mu\nu}^a(0)F_{\mu\nu}^a(0)\right>\right)^
{\frac{1}{2}}T_g^2$ and $\left(\frac{T_g}{r}\right)^2$, where
$T_g$ is the correlation length of the vacuum ($T_g\simeq 
0,13$ fm in the $SU(2)$ case$^{5}$, which will be studied below, and 
$T_g\simeq 0,22$ fm in the $SU(3)$ case$^{6}$, studied in Ref. 1)  
and $r\simeq 1$ fm is the size of the Wilson loop in the confining
regime$^{7}$. This yields the so-called curvature expansion for the
gluodynamics string effective action. In Ref. 1 only the first
nonvanishing terms of this expansion, corresponding to the lowest 
(second) order in the first parameter were accounted for, which
corresponds to the bilocal approximation$^{3,8-12}$, and then 
the expansion up to terms of the third order in the second 
parameter was elaborated out. The first two terms of this expansion
read as follows

$$S_{biloc.}=\sigma\int d^2\xi\sqrt{\hat g}+\frac{1}{\alpha_0}\int
d^2\xi\sqrt{\hat g}\hat g^{ij}\left(\partial_it_{\mu\nu}\right)
\left(\partial_j
t_{\mu\nu}\right), \eqno (1)$$
where

$$\sigma=4T_g^2\int d^2zD\left(z^2\right) \eqno (2)$$
is the string tension of the Nambu-Goto term, and

$$\frac{1}{\alpha_0}=\frac{1}{4}T_g^4\int d^2zz^2\left(2D_1\left(z^2
\right)-D\left(z^2\right)\right) \eqno (3)$$
is an inverse bare coupling constant of the rigidity term, while the
terms of the third order in $\left(\frac{T_g}{r}\right)^2$ contain
higher derivatives of the induced metric $\hat g_{ij}$ and/or of the
extrinsic curvature tensor $t_{\mu\nu}$ w.r.t. world sheet
coordinates, and we shall not quote them here (see Ref. 1 for the
details)\footnote{From now on, we shall use for the world sheet
indices the letters from the middle of the Latin alphabet, $i, j,
k,...,$ in order not to confuse them with the colour indices $a, b,
c,...$.}. In Eqs. (2) and (3) $D$ and $D_1$ stand
for two renormalization group invariant coefficient functions,
parametrizing the gauge-invariant bilocal correlator of gluonic field
strength tensors$^{3,8-11}$, and it is worth noting that since the
nonperturbative parts of these functions are related to each other as
$\left| D_1\right|\simeq\frac{1}{3}D$ according to lattice data$^{6}$, the
rigidity term inverse coupling constant (3) is negative, which
due to Ref. 13 agrees with the mechanism of confinement, based
on the dual Meissner effect$^{14}$. This
result confirms that the method of vacuum correlators, developed in
Refs. 3 and 8-12 (see also Refs. therein), provides us with the
consistent description of the confining gluodynamics vacuum. The
approach suggested in Ref. 1 enables one to express all the coupling
constants of the terms emerging in the string action in higher orders
of the curvature expansion through the gauge-invariant correlators of the 
gluonic field strength tensor only.

Notice also, that in Ref. 15 the action (1) was applied to the derivation
of the correction to the Hamiltonian of the QCD string with quarks$^{16}$ 
due to the rigidity term.

However, it should be emphasized that the curvature expansion
describes only the pure nonperturbative content of the gluodynamics
string theory. As it was argued in Ref. 11, in order to get the 
exponential 
growth of the multiplicity of states in the spectrum of the open bosonic 
string, one must account for the perturbative gluons
interacting with the string world sheet, which can be done in the 
framework of
the perturbation theory in the nonperturbative QCD vacuum$^{9,10}$.

In this Letter, we shall take this interaction into account in the
lowest order of perturbation theory and obtain the corresponding
correction to the action (1). To this end, one needs to integrate
over perturbative fluctuations in the expression for the Wilson loop 
average written through the non-Abelian Stokes theorem. This
procedure, however, looks rather difficult to be elaborated out in
the case when one makes use of the version of the non-Abelian Stokes
theorem, suggested in Refs. 2 and 3, due to the path-ordering, which 
remains in the expression for the Wilson loop after rewriting it
as a surface integral. In order to get rid of it, we
find it convenient to exploit in the following another version of the 
non-Abelian Stokes theorem, 
which was proposed in Ref. 17, where the path-ordering was replaced
by the integration over an auxiliary field from the
$SU\left(N_c\right)/\left[U(1)\right]^{N_c-1}$ coset space. For
simplicity, we shall consider the $SU(2)$-case, when this field is a
unit three-vector $\vec n$, which characterizes the instant
orientation in the colour space, and the non-Abelian Stokes theorem
takes a remarkably simple form. Integration over perturbative
fluctuations then leads to an interaction of the elements of the
string world sheet. This finally yields a correction to the 
rigidity term, while the
string tension of the Nambu-Goto term is not changed keeping 
its pure nonperturbative value (2). All the
points, mentioned above, will be worked out in the next Section.

The main results of the Letter are summarized in the Conclusion.

\vspace{6mm}
{\large \bf 2. An Action of the Gluodynamics String including
Perturbative Gluonic Contributions}

\vspace{3mm}
The starting point of the effective string theory, we are going
to derive, is the Wilson loop average in the $SU(2)$ gluodynamics $\left<
W(C)\right>=\left<tr{\,}P{\,}\exp\left(ig\oint\limits_C^{} dx_\mu
A_\mu^aT^a\right)\right>$. Rewriting it as a surface
integral by virtue of the non-Abelian Stokes theorem, suggested in
Ref. 17, splitting the total field $A_\mu^a$ into a strong nonperturbative 
background
$B_\mu^a$ ensuring confinement and perturbative fluctuations $a_\mu^a,
A_\mu^a=B_\mu^a+ a_\mu^a$, with $gB_\mu^a=O\left(1\right), ga_\mu^a=O
\left(g\right)$, 
and making use of the background field
formalism$^{9,10,18}$, we obtain

$$\left<W(C)\right>=N\int DB_\mu^a \eta\left(B_\alpha^b\right) 
Da_\mu^a D\vec n \exp\Biggl[\int dx
\Biggl[-\frac{1}{4}\left(F_{\mu\nu}^a\right)^2+\frac{1}{2}a_\mu^a
D_\lambda^{ab}D_\lambda^{bc}a_\mu^c+a_\nu^aD_\mu^{ab}F_{\mu\nu}^b\Biggr]+$$

$$+\frac{iJ}{2}\int
d\sigma_{\mu\nu}n^a\Biggl[ -g\left(F_{\mu\nu}^a+2D_\mu^{ab}a_\nu^b\right)+
\varepsilon^{abc}\Biggl[\left(D_\mu\vec n{\,}\right)^b\left(D_\nu\vec n{\,}
\right)^c-$$

$$-gn^d\left(\varepsilon^{bde}a_\mu^e\left(D_\nu\vec n{\,}\right)^c+
\varepsilon^{cde}a_\nu^e\left(D_\mu\vec n{\,}\right)^b\right)\Biggr]\Biggr]
\Biggr]. \eqno (4)$$
Here $F_{\mu\nu}^a=\partial_\mu B_\nu^a-\partial_\nu
B_\mu^a+g\varepsilon^{abc}B_\mu^bB_\nu^c$ is the strength tensor of the
background field,
$D_\mu^{ab}=\delta^{ab}\partial_\mu-g\varepsilon^{abc}B_\mu^c$ is the
corresponding covariant derivative, 
$J=\frac{1}{2},1,\frac{3}{2},...$ is the colour ``spin'' of the 
representation
of the Wilson loop under consideration, i.e. $T^aT^a=J(J+1)$, and the last 
term on the R.H.S. of Eq. (4) may be rewritten as follows

$$\exp\left[-\frac{iJg}{2}\int d\sigma_{\mu\nu}\varepsilon^{abc}n^a n^d 
\left(\varepsilon^{bde}a_\mu^e\left(D_\nu\vec n{\,}\right)^c +
\varepsilon^{cde}
a_\nu^e\left(D_\mu\vec n{\,}\right)^b\right)\right]=$$

$$=\exp\left[iJg\int d\sigma_{\mu\nu}a_\nu^a
\left[n^an^b\left(D_\mu\vec n{\,}
\right)^b-\left(D_\mu\vec n{\,}\right)^a\right]\right]=$$

$$=\exp\left[-iJg\int d\sigma_{\mu\nu} a_\nu^a\left(D_\mu\vec n{\,}
\right)^a\right],$$
where in the last line we have used the fact that $\vec n{\,}^2=1$, and 
$\varepsilon^{bcd}n^bn^c=0$.

It is worth noting, that the averaging over the fields $B_\mu^a$ 
and $a_\mu^a$ in Eq. (4) has been performed separately by making 
use of the so-called 
't Hooft identity$^{10}$ valid for an arbitrary functional $f$

$$\int DA_\mu^a f\left(A_\alpha^b\right)=\frac{\int DB_\mu^a \eta
\left(B_\alpha^b\right)\int Da_\mu^a f\left(B_\alpha^b+a_\alpha^b\right)}
{\int DB_\mu^a\eta\left(B_\alpha^b\right)},$$
which enables one to avoid double counting of fields during the 
integration. Here an arbitrary weight $\eta\left(B_\alpha^b\right)$ 
should be fixed by the requirement that all the 
cumulants and the string tension of the Nambu-Goto term acquire 
their observed values. 

Let us discuss the approximations, which have been done during the 
derivation of Eq. (4). 
In what follows, we shall be
interested only in the effects of the lowest order of perturbation
theory, so that on the R.H.S. of Eq. (4) we have omitted the
ghost term, the terms describing the interaction of two perturbative 
gluons with the string 
world sheet, containing one or three $\vec n$-fields, and the terms 
which describe the self-interaction of three
and four perturbative gluons.  
Secondly, for simplicity we neglect the 
interaction of two perturbative gluons with the background field strength 
tensor (gluon spin interaction)\footnote{ Within the Feynman-Schwinger 
proper time representation for the perturbative gluon propagator (see Eq. 
(5) below), such a term leads to insertions of the colour magnetic moment 
into the contour of integration.}.    

Integration over the perturbative
fluctuations in Eq. (4) is then Gaussian and yields

$$\left<W(C)\right>=N\int DB_\mu^a \eta\left(B_\alpha^b\right) 
D\vec n \exp\Biggl[-\frac{1}{4}\int
dx\left(F_{\mu\nu}^a\right)^2+\frac{iJ}{2}\int
d\sigma_{\mu\nu}\varepsilon^{abc}n^a\left(D_\mu\vec n{\,}\right)^b
\left(D_\nu\vec
n{\,}\right)^c\Biggr]\cdot$$

$$\cdot\exp\Biggl(-\frac{iJg}{2}\int
d\sigma_{\mu\nu}n^aF_{\mu\nu}^a\Biggr)\cdot$$

$$\cdot\exp\Biggl[-\frac{1}{2}\int dxdy
\Biggl[-iJgT_{\mu\nu}(x)D_\mu^{ba}n^a(x) +D_\mu^{ba}
\left(iJgT_{\mu\nu}(x)n^a(x)+F_{\mu\nu}^a(x)\right)\Biggr]\cdot$$

$$\cdot\int\limits_0^\infty ds\int (Dz)_{xy}e^{-\int\limits_0^s
\frac{\dot
z^2}{4}d\lambda}\Biggl[P{\,}\exp\Biggl(ig\int\limits_0^s
d\lambda\dot z_\alpha
B_\alpha\Biggr)\Biggr]^{bc}\cdot$$

$$\cdot\Biggl[-iJgT_{\rho\nu}(y)D_\rho^{cd}
n^d(y)+D_\rho^{cd}\left(
iJgT_{\rho\nu}(y)n^d(y)+F_{\rho\nu}^d (y)\right)\Biggr]\Biggr], \eqno
(5)$$
where $T_{\mu\nu}(x)\equiv\int
d^2\xi\varepsilon^{ij}\left(\partial_ix_\mu(\xi)\right)
\left(\partial_jx_\nu(\xi)\right)
\delta\left(x-x(\xi)\right)$ is the vorticity tensor current. 
In the bilocal approximation, we get up to an unimportant additive 
constant the following correction to 
the string
effective action (1)

$$\Delta S=-\frac{J^2g^2}{2}\int
dxdy\Biggl<(\partial_\mu T_{\mu\nu}(x))n^b(x)\int\limits_0^\infty
ds\int\left(Dz\right)_{xy}e^{-\int\limits_0^s\frac{\dot
z^2}{4}d\lambda}\cdot$$

$$\cdot\Biggl[P{\,}\exp\Biggl(ig\int\limits_0^s
d\lambda\dot z_\alpha
B_\alpha\Biggr)\Biggr]^{bc}(\partial_\rho T_{\rho\nu}(y))
n^c(y)\Biggr>_{\vec
n{\,}, B_\mu^a}, \eqno (6)$$
where

$$\left<...\right>_{\vec n{\,}}\equiv\int D\vec
n{\,}\left(...\right)\exp\left(\frac{iJ}{2}\int
d\sigma_{\mu\nu}\varepsilon^{abc}n^a\left(D_\mu\vec
n{\,}\right)^b\left(D_\nu \vec n{\,}\right)^c\right),$$

$$\left<...\right>_{B_\mu^a}\equiv\int
DB_\mu^a\left(...\right)\eta\left(B_\alpha^b\right)
\exp\left(-\frac{1}{4}\int
dx\left(F_{\mu\nu}^a\right)^2\right),$$
and in the derivation of
Eq. (6) we have neglected 
the interaction of the string world sheet with the background 
sources of the type $D_\alpha^{ab}F_{\alpha\beta}^b(u)$, where $u$ is 
an arbitrary space-time point outside the world sheet, which should 
be finally integrated over.

Integrating in Eq. (6) by parts, we arrive at the following formula

$$\Delta S=-\frac{J^2g^2}{2}\int d^2\xi\int
d^2\xi^\prime\varepsilon^{ij}\varepsilon^{kl}\left(\partial_ix_\mu\right)
\left(\partial_jx_\nu\right)(\partial_kx_\rho^\prime)\left(
\partial_lx_\nu^\prime\right)\cdot$$

$$\cdot\frac{\partial^2}{\partial x_\mu\partial x_\rho^\prime}
\left<\left<n^b\left(x\right)n^c\left(
x^\prime \right)\right>_{\vec n{\,}}\int\limits_0^\infty
ds\int\left(Dz\right)_{xx^\prime}e^{-\int\limits_0^s\frac{\dot
z^2}{4}d\lambda}\left[P{\,}\exp\left(ig\int\limits_0^s d\lambda\dot
z_\alpha B_\alpha\right)\right]^{bc}\right>_{B_\mu^a}, \eqno (7)$$
where $x_\mu\equiv x_\mu\left(\xi\right)$, and $x_\mu^\prime\equiv
x_\mu\left(\xi^\prime\right)$.

Since $e^{-\int\limits_0^s\frac{\dot
z^2}{4}d\lambda}P{\,}\exp\left(ig\int\limits_0^s d\lambda\dot z_\alpha
B_\alpha\right)$ is the statistical weight of a perturbative gluon,
propagating from the point $x^\prime$ to the point $x$ along the
trajectory $z_\alpha$ during the proper time $s$, it is the region
where $s$ is small, which mainly contributes to the path integral on
the R.H.S. of Eq. (7). This means that the dominant contribution to
$\Delta S$ comes from those $x_\mu$'s and $x_\mu^\prime$'s, which are
very close to each other, which is in the line with the curvature
expansion, where $\left| x^\prime-x\right| \le T_g\ll r$. Within this
approximation, one gets

$$\left<n^b\left(x\right)n^c\left(x^\prime\right)\right>_{\vec n{\,}}
\simeq\frac{\delta^{bc}}{3}\int D\vec n{\,}\exp\left(\frac{iJ}{2}
\int d\sigma_{\mu\nu}\varepsilon^{def}n^d\left(D_\mu\vec
n{\,}\right)^e\left(D_\nu\vec n{\,}\right)^f\right). \eqno (8)$$

It is worth noting, that the integral on the R.H.S. of Eq. (8) 
is a functional of the world sheet as a whole (it is independent of 
$x_\mu(\xi)$), and therefore may be absorbed into the integration 
measure.  

Hence, as it was announced in the Introduction, we see that
expression (7) for the correction to the string effective action
(1) due to the perturbative gluons takes the form of the interaction
of two elements of the world sheet, $d\sigma_{\mu\nu}(\xi)$ and
$d\sigma_{\rho\nu}\left(\xi^\prime\right)$, via the nonperturbative
gluonic string, which emerges from the perturbative gluonic exchanges 
between these world sheet elements.

For performing the derivatives of the $B_\mu^a$-average in Eq. (7), 
let us extract explicitly the dependence on the points $x$ and
$x^\prime$ from the functional integral. 
To this end, it is convenient to pass to the integration 
over the trajectories
$u_\mu(\lambda)=z_\mu(\lambda)+\frac{\lambda}{s}\left(x^\prime-
x\right)_\mu - x_\mu^\prime$, which yields

$$\Delta S=-\frac{J^2g^2}{6}\int d^2\xi\int
d^2\xi^\prime\varepsilon^{ij}\varepsilon^{kl}\left(\partial_ix_\mu\right)
\left(\partial_j x_\nu\right)(\partial_k
x_\rho^\prime)\left(\partial_lx_\nu^\prime\right)\frac{\partial^2}{\partial 
x_\mu\partial x_\rho^\prime}\int\limits_0^\infty ds e^{-\frac{\left(x-
x^\prime\right)^2}{4s}}\cdot$$

$$\cdot \int \left(Du\right)_{00}e^{-\int\limits_0^s\frac
{\dot u^2}{4}d\lambda}
\left<tr{\,}P{\,}\exp\left[ig\int\limits_0^s d\lambda
\left(\frac{x-x^\prime}{s}+\dot u\right)_\alpha B_\alpha\left(u+x^\prime+ 
\frac{\lambda}{s}\left(x-x^\prime\right)\right)\right]\right>_{B_\mu^a}. 
\eqno (9)$$

In the following, we shall be interested only 
in the contributions emerging from 
Eq. (9) to the Nambu-Goto and rigidity terms. As it turns out, there exist 
two origins of such contributions. The first of them arises from taking 
the derivatives of the factor $e^{-\frac{\left(x-x^\prime\right)^2}{4s}}$ 
and replacing the Wilson loop average by $\left<tr{\,}P{\,}\exp\left(
ig\int\limits_0^s d\lambda\dot u_\alpha 
B_\alpha (u)\right)\right>_{B_\mu^a}.$  
The second one is due to acting with the derivatives upon the Wilson loop 
average and leads to the correlator of the $B_\mu^a$-fields. For the case 
of a fastly converging cumulant expansion when $\left(\alpha_s\left<
F_{\mu\nu}^a(0)F_{\mu\nu}^a(0)\right>\right)^{\frac{1}{2}}T_g^2\ll\left(
\frac{T_g}{r}\right)^2$, one can neglect the second contribution w.r.t. 
the first one. Thus we obtain   

$$\Delta S=\frac{J^2g^2}{12}\int d^2\xi\int d^2\xi^\prime
\varepsilon^{ij}\varepsilon^
{kl}(\partial_ix_\mu)(\partial_jx_\nu)(\partial_kx_\rho^\prime)(\partial_l
x_\nu^\prime)\cdot$$

$$\cdot\int\limits_0^\infty\frac{ds}{s}\left(\frac{\left(x-x^\prime
\right)_\mu\left(x-x^\prime\right)_\rho}{2s}-\delta_{\mu\rho}\right)
e^{-\frac{\left(x-x^\prime\right)^2}{4s}}\Phi (s), \eqno (10)$$
where

$$\Phi (s)\equiv\int \left(Du\right)_{00}e^{-\int\limits_0^s\frac{\dot 
u^2}{4}d\lambda}\left<tr{\,}P{\,}\exp\left(ig\int\limits_0^s d\lambda
\dot u_\alpha B_\alpha(u)\right)\right>_{B_\mu^a}. \eqno (11)$$

Finally, in order to get the desirable correction to the action (1),
we shall Taylor expand the R.H.S. of Eq. (10) in powers of $\frac{s}{r^2}$
(according to the discussion in the paragraph before Eq. (8)),
keeping in this expansion terms not higher in the derivatives w.r.t.
world sheet coordinates than in the rigidity term, which corresponds to the
expansion up to the second order in the parameter $\frac{s}{r^2}$. To this 
end, we shall make use of the Gauss-Weingarten formulae 

$$D_iD_jx_\mu=
\partial_i\partial_jx_\mu-\Gamma_{ij}^k\partial_kx_\mu=K_{ij}^IN_\mu^I,{\,}
N_\mu^I N_\mu^{I^\prime}=
\delta^{II^\prime}, {\,}N_\mu^I \partial_ix_\mu=0,{\,}I,I^\prime=1,2,$$ 
where $\Gamma_{ij}^k$ is a Christoffel symbol, $K_{ij}^I$ is the second 
fundamental form of the string world sheet, and $N_\mu^I$'s are the unit 
normals to the sheet. 
Omitting the full derivative terms of the form $\int
d^2\xi\sqrt{\hat g}R$, where $R$ is a scalar curvature of the world sheet,
we get analogously to Ref. 1 the following values of the integrals 
standing on the R.H.S. of Eq. (10)

$$\int d^2\xi\int d^2\xi^\prime \varepsilon^{ij}\varepsilon^{kl}
\left(\partial_ix_\mu\right)\left(\partial_jx_\nu\right)(\partial_k
x_\rho^\prime)\left(\partial_lx_\nu^\prime\right)
\left(x-x^\prime\right)_\mu
\left(x-x^\prime\right)_\rho\int\limits_0^\infty\frac{ds}{s^2}
e^{-\frac{\left(x-x^\prime\right)^2}{4s}}\Phi
(s)=$$

$$=4\pi\int\limits_0^\infty ds\Phi (s)\left(4\int
d^2\xi\sqrt{\hat g}-3s\int d^2\xi\sqrt{\hat g}\hat g^{ij}\left(\partial_i
t_{\mu\nu}\right)\left(\partial_j t_{\mu\nu}\right)\right) \eqno
(12)$$
and

$$\int d^2\xi\int d^2\xi^\prime\varepsilon^{ij}\varepsilon^{kl}\left(
\partial_ix_\mu\right)\left(\partial_jx_\nu\right)(\partial_k
x_\mu^\prime)\left(\partial_lx_\nu^\prime\right)\int\limits_0^\infty
\frac{ds}{s}e^{-\frac{\left(x-x^\prime\right)^2}{4s}}\Phi
(s)=$$

$$=2\pi\int\limits_0^\infty ds\Phi (s)\left(4\int
d^2\xi\sqrt{\hat g}-s\int d^2\xi\sqrt{\hat g}\hat g^{ij}\left(\partial_i
t_{\mu\nu}\right)\left(\partial_j t_{\mu\nu}\right)\right). \eqno
(13)$$
Combining Eqs. (12) and (13) together, we arrive at
the following correction to the effective action (1) due to the
accounting for the perturbative gluons in the lowest order of
perturbation theory

$$\Delta S=\left(\Delta\frac{1}{\alpha_0}\right)\int
d^2\xi\sqrt{\hat g}\hat g^{ij}\left(\partial_i
t_{\mu\nu}\right)\left(\partial_j t_{\mu\nu}\right), \eqno (14)$$
where

$$\left(\Delta\frac{1}{\alpha_0}\right)=-\frac{\pi}{3}J^2g^2\int
\limits_0^\infty
dss\Phi (s). \eqno (15)$$
Notice, that as it was already pointed out in the Introduction,
perturbative gluons do not change the value of the string tension (2)
of the Nambu-Goto term and affect only the coupling constant of the
rigidity term. The nontrivial content of correction (15) to the 
nonperturbative rigidity coupling constant (3) emerges due to the 
path integral defined by Eq. (11), which includes the background Wilson 
loop average. Since this correction is a pure perturbative effect, its
sign, which depends on the background entering Eq. (11), 
is unimportant for the explanation of confinement in terms of
the dual Meissner effect (see discussion in the Introduction).

\vspace{6mm}
{\large \bf 3. Conclusion}

\vspace{3mm}
In this Letter, we have applied perturbation theory in the
nonperturbative QCD vacuum$^{9,10}$ and the non-Abelian Stokes theorem,
which represents a Wilson loop in the $SU(2)$ gluodynamics as an
integral over all the orientations in colour space$^{17}$ to the
derivation of the correction to the string effective action (1), found in
Ref. 1. The resulting perturbative gluonic correction is given by formulae
(11), (14) and (15) and affects only the rigidity term, while the string
tension of the Nambu-Goto term keeps its pure nonperturbative value
(2). The perturbative contribution to the inverse coupling constant of
the rigidity term contains the dependence on the background fields in
the form of the background Wilson loop average standing under a certain
path integral (11).

We have also demonstrated that perturbative fluctuations, when being
taken into account, lead to the interaction defined by the R.H.S. of
Eq. (7) between elements of the string world sheet by virtue of
nonperturbative gluonic strings, which agrees with the qualitative
scenario of the excitation of the gluodynamics string by the perturbative
gluons, suggested in Refs. 1 and 11.

However, it still remains unclear whether perturbative gluons may
yield a cancellation of the conformal anomaly in $D=4$ rather than in
$D=26$, as it takes place for the ordinary bosonic string 
theory$^{19,20}$, and the
solution of the problem of crumpling for the rigidity term$^{20,21}$. 
There also exist some other obstacles with the rigidity term, such as the 
absence of the lowest-energy state in this theory and the wrong 
high-temperature 
behaviour of the free energy per unit length of the string. These 
difficulties are absent in the recently proposed nonlocal theory of a string 
with the negative stiffness$^{22}$. At the moment, we do not know how 
this theory could be obtained in the framework of our approach. 
The problems mentioned above will be the topics of the next publications.

\vspace{6mm}
{\large \bf 4. Acknowledgments}

\vspace{3mm}
We are deeply grateful to H.Dorn, H.Kleinert, Yu.M.Makeenko, 
M.Mueller-Preussker, Chr.Preitschopf, Chr.Schubert, and Yu.A.Simonov 
for a lot of useful
discussions on the problem studied in this paper. One of us (D.A.) 
would also like
to thank the theory group of the Quantum Field Theory Department of
the Institut f\"ur Physik of the Humboldt-Universit\"at of Berlin
for kind hospitality. 

\vspace{6mm}
{\large \bf References}

\vspace{3mm}
\noindent
1.~D.V.Antonov, D.Ebert and Yu.A.Simonov, {\it Mod.Phys.Lett.} 
{\bf A11}, 1905 (1996) (preprint DESY 96-134).\\
2.~M.B.Halpern, {\it Phys.Rev.} {\bf D19}, 517 (1979); I.Ya.Aref'eva, 
{\it Theor.Math.Phys.} {\bf 43}, 111 (1980); N.Brali\'c, {\it Phys.Rev.} 
{\bf D22}, 3090 (1980).\\
3.~Yu.A.Simonov, {\it Yad.Fiz.} {\bf 50}, 213 (1989).\\
4.~N.G. Van Kampen, {\it Stochastic Processes in Physics and
Chemistry} (North-Holland Physics Publishing, 1984).\\
5.~M.Campostrini, A. Di Giacomo and G.Mussardo, {\it Z.Phys.} 
{\bf C25}, 173 (1984).\\
6.~A. Di Giacomo and H.Panagopoulos, 
{\it Phys.Lett.} {\bf B285}, 133 (1992).\\
7.~I.-J.Ford et al., {\it Phys.Lett.} {\bf B208}, 286 (1988); 
E.Laermann et al., {\it Nucl.Phys.} {\bf B26} (Proc.Suppl.), 268 (1992).\\
8.~H.G.Dosch, {\it Phys.Lett.} {\bf B190}, 177 (1987); Yu.A.Simonov, 
{\it Nucl.Phys.} {\bf B307}, 512 (1988); H.G.Dosch and Yu.A.Simonov, 
{\it Phys.Lett.} {\bf B205}, 339 (1988), {\it Z.Phys.} {\bf C45}, 147 
(1989); Yu.A.Simonov, {\it Nucl.Phys.} {\bf B324}, 67 (1989), 
{\it Phys.Lett.} {\bf B226}, 151 (1989), {\it Phys.Lett.} {\bf B228}, 413 
(1989), {\it Yad.Fiz.} {\bf 54}, 192 (1991); H.G.Dosch, A. Di Giacomo 
and Yu.A.Simonov, in preparation.\\
9.~Yu.A.Simonov, {\it Yad.Fiz.} {\bf 58}, 113, 357 (1995), preprint 
ITEP 37-95; D.V.Antonov, {\it Phys.Atom.Nucl.} {\bf 60}, 299 (1997) 
({\it hep-th}/9605044).\\
10. E.L.Gubankova and Yu.A.Simonov, {\it Phys.Lett.} 
{\bf B360}, 93 (1995); Yu.A.Simonov, in {\it Lecture Notes in Physics}, 
v. 479 (Springer Verlag, 1997); A.M.Badalian 
and Yu.A.Simonov, {\it Yad.Fiz.} {\bf 60}, 714 (1997).\\ 
11. Yu.A.Simonov, {\it Nuovo Cim.} {\bf A107}, 2629 (1994).\\
12. D.V.Antonov and Yu.A.Simonov, {\it Int.J.Mod.Phys.} {\bf A11}, 4401 
(1996); D.V.Antonov, {\it JETP Lett.} {\bf 63}, 398 (1996), 
{\it Mod.Phys.Lett.} {\bf A11}, 3113 (1996) ({\it hep-th}/9612005), 
{\it Int.J.Mod.Phys.} {\bf A12}, 2047 (1997), {\it Phys.Atom.Nucl.} 
{\bf 60}, 478  
(1997) ({\it hep-th}/9605045).\\
13. P.Orland, {\it Nucl.Phys.} {\bf B428}, 221 (1994).\\
14. S.Mandelstam, {\it Phys.Lett.} {\bf B53}, 476 (1975); G.'t Hooft, 
in {\it High Energy Physics}, Ed. A.Zichichi (Editrice Compositori, 
1976).\\
15. D.V.Antonov, {\it JETP Lett.} {\bf 65}, 701 (1997) 
({\it hep-th}/9612109).\\
16. A.Yu.Dubin, A.B.Kaidalov and Yu.A.Simonov, {\it Yad.Fiz.} 
{\bf 56}, 213 (1993), {\it Phys.Lett.} {\bf B323}, 41 (1994); 
E.L.Gubankova and A.Yu.Dubin, {\it Phys.Lett.} {\bf B334}, 180  
(1994), preprint ITEP 62-94.\\
17. D.I.Diakonov and V.Yu.Petrov, in {\it Nonperturbative Approaches
to QCD}, Proceedings of the International Workshop at ECT*, Trento, 
July 10-29, 1995, Ed. D.I.Diakonov (PNPI, 1995), 
{\it hep-th}/9606104.\\
18. B.S. De Witt, {\it Phys.Rev.} {\bf 162}, 1195, 1239 (1967); 
J.Honerkamp, {\it Nucl.Phys.} {\bf B48}, 269 (1972); G.'t Hooft, 
{\it Nucl.Phys.} {\bf 62}, 444 (1973); L.F. Abbot, {\it Nucl.Phys.} 
{\bf B185}, 189 (1981).\\
19. A.M.Polyakov, {\it Phys.Lett.} {\bf B103}, 207 (1981).\\
20. A.M.Polyakov, {\it Gauge Fields and Strings} (Harwood Academic 
Publishers, 1987).\\
21. A.M.Polyakov, {\it Nucl.Phys.} {\bf B268}, 406 (1986).\\
22. H.Kleinert and A.M.Chervyakov, {\it hep-th}/9601030 (in press in 
{\it Phys.Lett.} {\bf B}).
\end{document}